\documentclass[showpacs,prl,twocolumn,amsmath,amssymb]{revtex4-1} %
\usepackage{epsfig,amsmath}
\usepackage{subfigure}
\usepackage{graphicx}% Include figure files
\usepackage{dcolumn}% Align table columns on decimal point
\usepackage{stmaryrd}
\usepackage{mathrsfs}
\usepackage{pifont}
\usepackage{amsthm}
\usepackage{amssymb}
\usepackage{bm}
\usepackage{latexsym}

\newcommand{\la}{\langle}
\newcommand{\ra}{\rangle}
\newcommand{\beq}{\begin{equation}}
\newcommand{\eeq}{\end{equation}}
\newcommand{\beqa}{\begin{eqnarray}}
\newcommand{\eeqa}{\end{eqnarray}}
 
\newcommand{\ti}{\tilde} 
 
\newcommand{\da}{\dagger} 
 
 \newcommand{\om}{\omega}
 
\newcommand{\non}{\nonumber}

\def\jpa#1{{ J.\ Phys.\ A} {\bf#1}}
\def\pra#1{{ Phys.\ Rev. A\/} {\bf#1}}
\def\prb#1{{ Phys.\ Rev. B\/} {\bf#1}}

\def\prl#1{{ Phys.\ Rev.\ Lett.} {\bf#1}}

\def\annph#1{{ Ann.\ Phys.} {\bf #1}}
\def\pla#1{{ Phys.\ Lett. A\/} {\bf#1}}
\def\rmp#1{{ Rev. \ Mod. \ Phys.} {\bf#1}}

\makeatletter
\renewcommand*\@fnsymbol[1]{\the#1}
\makeatother

\begin{document}

\title{Non-Markovian Relaxation of a Three-Level System:  Quantum
Trajectory Approach}

\author{Jun Jing}\email{Email address:Jun.Jing@stevens.edu}
\altaffiliation[Also at ]{Department of Physics, Shanghai University,
Shanghai 200444, China}

\author{Ting Yu}\email{Email address:Ting.Yu@stevens.edu}

\affiliation{Center for Controlled Quantum Systems and
Department of Physics and Engineering Physics, Stevens Institute of
Technology, Hoboken, New Jersey 07030, USA}

\date{\today}

\begin{abstract}
The non-Markovian dynamics of a three-level quantum system coupled to a bosonic
environment is a difficult problem due to the lack of an exact dynamic equation
such as a master equation.  We present for the first time an exact quantum
trajectory approach to a dissipative three-level model.  We have established a
convolutionless stochastic Schr\"{o}dinger equation called time-local quantum
state diffusion (QSD) equation without any approximations, in particular,
without Markov approximation. Our exact time-local QSD equation opens a new
avenue for exploring quantum dynamics for a higher dimensional quantum system
coupled to a non-Markovian environment.
\end{abstract}

\pacs{03.65.-w, 05.30.-d, 42.50.Lc}

\maketitle

Every small quantum system, such as a two-level atom (qubit), a three-level
atom (qutrit) or a cavity mode (simple harmonic oscillator), should be regarded
as an open quantum system due to the inevitable interaction with its
environment. The effect from the environment may bring about some grave
problems, such as quantum decoherence and disentanglement, to the dynamics of
the open system \cite{Breuer}. Much has been studied in the Markov limit where
the dynamics of open quantum system is typically described by a standard
Lindblad master equation which is equivalent to a quantum state diffusion (QSD)
equation for a pure state
\cite{Carmichael,Dalibardetal93,Gisin-Percival93,Gardiner,Wiseman,Knight}
(setting $\hbar=1$):
\begin{equation}\label{QSD}
\frac{d}{dt}\psi_t=-iH\psi_t+\Delta_t(L)(z_t+\la
L^\da\ra_t)\psi_t-\frac{1}{2}\Delta_t(L^\da L)\psi_t,
\end{equation}
where the notation $\Delta_t(A)\equiv A-\la A\ra_t$ for any operator $A$ and
$\la A\ra_t\equiv\la\psi_t|A|\psi_t\ra$ denotes the quantum average. $L$ is the
system operator coupled to the environment, often called Lindblad operator. The
dynamics of the open system is driven by the classical stochastic process
$z_t$. The reduced density matrix for the system of interest can be recovered
by averaging quantum trajectories generated by the QSD equation (\ref{QSD}):
$\rho_t=M[|\psi_t(z)\ra\la\psi_t(z)|]$. Here $M[\cdot]$ denotes the ensemble
average over the classical noise. Besides many appealing features exhibited by
pure state quantum trajectories, Eq.~(\ref{QSD}) provides a very efficient
numerical tool in solving quantum dynamics of Markov open systems.

Non-Markovian environments have become increasingly important in recent times
due to their relevance in explaining new experimental advances in high-Q
microwave cavities, photonic crystals and atom laser in BEC
\cite{Hope,John,Bay,Vats,Moy,Daniel,Franco}. It is also evident from the recent
progress in quantum information processing, that environmental memory may be
utilized to generate or modulate entanglement evolution of an open quantum
system \cite{non-Markovian1,non-Markovian2}. Clearly, an approach that is
capable of dealing with non-Markovian quantum system is highly desirable. A
non-Markovian QSD equation for a general non-Markovian open system by
Di\'{o}si, Gisin and Strunz has provided a powerful tool in dealing with the
exact dynamics of quantum open systems irrespective of the coupling strength,
the correlation time and the spectral density of the bosonic environments
\cite{Diosi97,Diosi,Strunz99,Yupra99,Strunzetal99,Strunz01CP,Wisemanetal2002,
nonMark}. Despite extensive research, the explicit non-Markovian QSD equations
only exist for a single two-level system, the quantum Brownian motion model and
optical cavities due to the intricate functional derivative appearing in the
fundamental QSD equation \cite{Diosi97,
Diosi,Feynman,Leggett,Hupazzhang,Wmzhang}. Clearly, the power of the
non-Markovian quantum trajectory method cannot be readily unleashed unless the
stochastic QSD equation can be cast into a convolutionless form in which the
numerical simulations and analytical applications can be easily implemented
\cite{SY,Yu}.

In this Letter, we present, for the very first time, an exact time-local QSD
equation for the three-level dissipative dynamics in the framework of
non-Markovian quantum trajectory at zero temperature. Our treatments of the
three-level model have opened up a new route of exploring dynamics of higher
dimensional quantum open systems. Notably, our method of establishing a
time-local non-Markovian QSD equation can be applied to multiple qubit systems
to deal with the time behavior of them such as the estimation of non-Markovian
entanglement evolution and fidelity \cite{JingandYu2010,carlo2003}.

We consider a Caldeira-Leggett-like model \cite{Leggett} involving a system
with Hamiltonian $H$ describing a three-level atom, coupled linearly to a
general boson environment consisting of a set of harmonic oscillators $a_{\bf
k}, a_{\bf k}^\dag$ (e.g., cavity modes). The total Hamiltonian for the system
of interest plus environment can be written as:
\begin{equation}\label{Hamil}
H_{\rm tot}=H+\sum_{\bf k}(g^*_{\bf k}L^\da a_{\bf k}+g_{\bf k}La_{\bf k}^\da)
+\sum_{\bf k}\omega_{\bf k}a_{\bf k}^\da a_{\bf k},
\end{equation}
where the three-level system Hamiltonian $H=\omega
J_z=\omega(|2\ra\la2|-|0\ra\la0|)$, the Lindblad operator
$L=J_-=\sqrt{2}(|0\ra\la1|+|1\ra\la2|)$ and $L^\da=J_+$. Note that $\omega$ is
the spacing of the two neighboring energy levels of the three-level system,
$g_{\bf k}$ are the coupling constants between the three-level system and the
environmental modes. It is known \cite{Diosi97} that the linear QSD equation
for the three-level system can be formally written as:
\begin{eqnarray}\non
\frac{d}{dt}\psi_t(z)&=&-i\om J_z\psi_t(z)+J_-z_t\psi_t(z)\\ \label{LQSD}
&-&J_+\int_0^tds\alpha(t,s)\frac{\delta\psi_t(z)}{\delta z_s},
\end{eqnarray}
where $\alpha(t,s)$ is the bath correlation function and $z_t=-i\sum_{\bf k}
g^*_{\bf k}z^*_{\bf k}e^{i\omega_{\bf k}t}$ is a complex Gaussian process
satisfying $M[z_t]=M[z_tz_s]=0$ and $M[z^*_tz_s]=\alpha(t,s)$. When
$\alpha(t,s)=\delta(t-s)$, it reduces to the memoryless Markov case. The
density matrix of the system is recovered from the ensemble average of many
realizations of quantum trajectories:
\begin{equation}\label{rhot}
\rho_t=M[|\psi_t(z)\ra\la\psi_t(z)|]=\int\frac{d^2z}{\pi}e^{-|z|^2}
|\psi_t(z)\ra\la\psi_t(z)|.
\end{equation}

One advantage of using quantum trajectory is that, while deriving non-Markovian
master equation is known to be a daunting task, the QSD equation governing the
non-Markovian open system can be read off directly once the Hamiltonian of the
total system is given by (\ref{Hamil}). However, the formal non-Markovian QSD
equation (\ref{LQSD}) or its nonlinear version (see below) cannot be easily
implemented as an analytical or numerical tool unless the term containing
functional derivative can be written as $ \frac{\delta}{\delta z_s}\psi_t(z)=
O(t,s,z)\psi_t(z)$, where $O(t,s,z)$ is an operator acting on the system
Hilbert space satisfying the initial condition $O(s,s,z)=J_-$. The equation of
motion governing the O-operator (consistency condition \cite{Diosi}) is given
by
\begin{eqnarray}\non
\partial_tO(t,s,z)&=&\left[-i\om J_z+J_-z_t-J_+\bar{O}(t,z),
O(t,s,z)\right] \\ &-& J_+\frac{\delta\bar{O}(t,z)}{\delta z_s},
\end{eqnarray}
where $\bar{O}(t,z)=\int^t_0 ds \alpha(t,s)O(t,s,z)$. For the model given by
(\ref{Hamil}), it can be shown that this nonlinear operator equation can be
solved by the following ansatz solution:
\begin{equation} \label{Oop}
O=f(t,s)J_-+g(t,s)J_zJ_-+i\int_0^tp(t,s,s')z_{s'}ds'J_-^2,
\end{equation}
where the coefficient functions $f(t,s)$, $g(t,s)$ and $p(t,s,s')$ satisfy:
\begin{eqnarray}\label{ft}
\frac{\partial}{\partial t}f(t,s)&=&i\omega
f(t,s)+2G(t)f(t,s)\non \\
&-&2iP(t,s) \\ \label{gt} \frac{\partial}{\partial t}g(t,s)&=&i\omega
g(t,s)-2F(t)f(t,s)\non \\
&+&2F(t)g(t,s) +4G(t)f(t,s)\non\\
&-&2G(t)g(t,s)-2iP(t,s) \\ \label{pts} \frac{\partial}{\partial
t}p(t,s,s')&=&2i\omega
p(t,s,s')+2F(t)p(t,s,s')\non\\
&+& 2f(t,s)P(t,s')-2g(t,s)P(t,s')
\end{eqnarray}
where the time-dependent functions $F(t)\equiv\int_0^t\alpha(t,s)f(t,s)ds$,
$G(t)\equiv\int_0^t\alpha(t,s)g(t,s)ds$ and
$P(t,s')\equiv\int_0^t\alpha(t,s)p(t,s,s')ds$ together with a set of boundary
conditions for $f, g$ and $p$: $f(s,s)=1$, $p(t,s,t)=-ig(t,s)$, $g(s,s)=0$,
$p(s,s,s')=0$.

With the explicit O-operator (\ref{Oop}), the exact linear non-Markovian QSD
equation may be compactly written into a time-local form, \beqa\label{LQSD3}
\frac{d}{dt}\psi_t(z) &=& \Big(-i\omega J_z+J_-z_t-F(t)J_+J_{-} -G(t)J_+
J_zJ_{-} \non \\ &-& i\int_0^tP(t,s')z_{s'}dsJ_+J_-^2\Big)\psi_t(z). \eeqa Note
that the time dependent coefficients $F(t)$, $G(t)$ and $P(t,s')$ could be
calculated once the correlation function $\alpha(t,s)$ is explicitly given. It
should be emphasized that the non-Markovian properties are encoded in a finite
width correlation function $\alpha(t,s)$, and so in the time-dependent
coefficients $F(t)$, $G(t)$ and $P(t,s')$ appearing in the QSD equation
(\ref{LQSD3}). Clearly, the terms containing $G(t)$ and $P(t,s')$ give rise to
the most important correction to the Markovian dynamics. This can be easily
seen if we take the Markov limit in which $F(t)=1/2, G(t)=P(t,s')=0$ recovering
the standard Markovian QSD equation. Eq.~(\ref{LQSD3}) is the fundamental
equation for the three-level system coupled to a bosonic environment described
by (\ref{Hamil}).

For numerical simulations, it is more efficient to use the nonlinear
non-Markovian QSD equation \cite{Diosi} for the normalized state
$\ti{\psi_t}=\frac{\psi_t}{||\psi_t||}$ obtained from Eq.~(\ref{LQSD3}) :
\begin{widetext}
\begin{eqnarray}
\frac{d}{dt}\ti{\psi}_t&=&-i\omega J_z\ti{\psi}_t
+\Delta_t(J_-)\ti{z}_t\ti{\psi}_t+\la J_{+}\ra_tF(t)\Delta_t(J_-)\ti{\psi}_t
-F(t)\Delta_t(J_+J_-)\ti{\psi}_t + \la J_{+}\ra_tG(t)\Delta_t(J_zJ_-)\ti{\psi}_t
\non\\ && -G(t)\Delta_t(J_+J_zJ_-)\ti{\psi}_t+ i\la
J_{+}\ra_t\int_0^tP(t,s')\ti{z}_{s'}ds'\Delta_t(J_-^2)\ti{\psi}_t-i\int_0^tP(t,
s')\ti{z}_{s'}ds'\Delta_t(J_+J_-^2)\ti{\psi}_t.
\label{NMQSD}
\end{eqnarray}
\end{widetext}
Here $\ti{z}_{t}=z_{t}+\int_0^{t}\alpha^*(t,s)\la J_+\ra_sds$ is the shifted
complex Gaussian process.

The correlation function $\alpha(t,s)$ at zero temperature is given by
$\alpha(t,s)=\sum_{\bf k}|g_{\bf k}|^2e^{-i\omega_{\bf k}(t-s)}=\int_0^\infty
d\omega S(\omega)e^{-i\omega(t-s)}$, where $S(\omega)$ is the spectral density
of the environment modes. The Eq.~(\ref{NMQSD}) is capable of describing
non-Markovian dynamics with an arbitrary correlation function $\alpha(t,s)$,
but for the sake of simplicity, we now consider correlation function
$\alpha(t,s)=\frac{\gamma}{2}e^{-\gamma|t-s|}$ for the Ornstein-Uhlenbeck
process. The Ornstein-Uhlenbeck process is a useful approach to modeling noisy
relaxation with a finite environmental memory time scale $1/\gamma$.  When
$\gamma\rightarrow\infty$, the environment memory time approaches to zero and
$\alpha(t,s)$ reduces to $\delta(t-s)$, which corresponds to the Markov limit.
Then Eq.~(\ref{NMQSD}) reduces to the standard nonlinear QSD equation
\cite{Gisin-Percival93}:
\begin{equation}
\frac{d}{dt}\ti{\psi}_t=-i\omega J_z\ti{\psi}_t+\Delta_t(J_-)(z_t+\la
J_+\ra_t)\ti{\psi}_t-\frac{1}{2}\Delta_t(J_+J_-)\ti{\psi}_t.
\end{equation}

For the Ornstein-Uhlenbeck correlation, the partial differential equations
(\ref{ft}), (\ref{gt}) and (\ref{pts}) could be converted into a set of
ordinary differential equations:
\begin{eqnarray}
\frac{dF(t)}{dt}&=&\frac{\gamma}{2}+(-\gamma+i\omega)F(t)\nonumber\\
&&+2F(t)G(t)-2i\bar{P}(t), \\
\frac{dG(t)}{dt}&=&(-\gamma+i\omega)
G(t)-2F^2(t)\nonumber \\
&&+6F(t)G(t)-2G^2(t)-2i\bar{P}(t), \\
\frac{d\bar{P}(t)}{dt}&=&-i\frac{\gamma}{2}G(t)+2(-\gamma+i\omega)
\bar{P}(t)\nonumber \\
&&+4F(t)\bar{P}(t)-2G(t)\bar{P}(t)
\end{eqnarray}
where $\bar{P}(t)=\int_0^t\alpha(t,s')P(t,s')ds'$ and the initial conditions are
given by $F(0)=G(0)=\bar{P}(0)=0$. It is easy to show that
$P(t,s')=-iG(s')e^{\int_{s'}^t[-\gamma+2i\omega+4F(s)-2G(s)]ds}$.

\begin{figure}[htbp]
\centering
\includegraphics[width=2.5in]{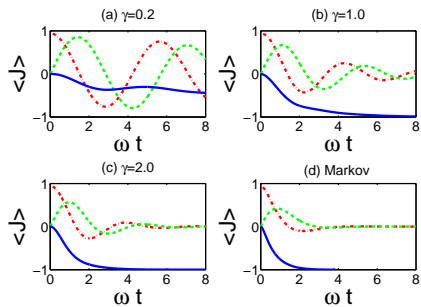}
\caption{Ensemble average of $\la\vec{J}\ra$ over $1000$ realizations (red
dot-dashed line for $\la J_x\ra$, green dashed line for $\la J_y\ra$ and blue
solid line for $\la J_z\ra$) with different $\gamma$s. Here we choose $\omega=1$
and the initial state $|\psi_0\ra=1/\sqrt{3}(|0\ra+|1\ra+|2\ra)$.} \label{JM}
\end{figure}

\begin{figure}[htbp]
\centering
\includegraphics[width=2.3in]{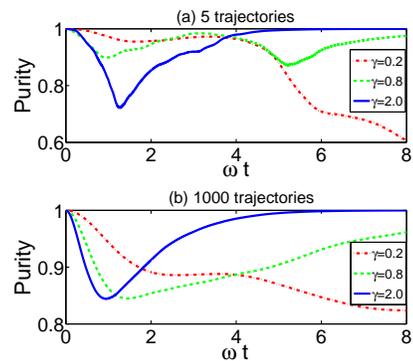}
\caption{Ensemble average of ${\rm Purity}=Tr[\rho^2(t)]$ over different number of
realizations, with red dot-dashed line for $\gamma=0.2$, green dashed line for
$\gamma=0.8$ and blue solid line for $\gamma=2.0$. Here we choose $\omega=1$ and
the initial state $|\psi_0\ra=1/\sqrt{3}(|0\ra+|1\ra+|2\ra)$.} \label{Pu}
\end{figure}

With the nonlinear non-Markovian QSD equation (\ref{NMQSD}), the simulations of
the three-level system (a spin-$1$ particle or a three-level atom) dynamics can
be efficiently implemented by realizing Gaussian sample paths. We first
calculated the ensemble average of the angular momentum time evolution with
different memory parameter $\gamma$. The plots are shown in Fig.~1. The
zero-temperature environment prohibits the transitions from a lower level to a
higher level, so for the three-level atom with an arbitrary initial state, the
spontaneous emission always causes the system to decay into its ground level
$|0\ra$ in the long time limit. Therefore, when $t\rightarrow\infty$, we get
$\la J_x\ra=\la J_y\ra\rightarrow0$ and $\la J_z\ra\rightarrow-1$, this scenario
can be easily seen from the Markov limit shown in Fig.~1(d). The non-Markovian
features of the environment for different $\gamma$ are illustrated in
Figs.~1(a), 1(b) and 1(c), where we can see that the transition of dynamics from
non-Markovian to Markov regimes is dictated by environment memory time
$\tau=1/\gamma$. Clearly, the non-Markovian features are lost when the system
approaches to its Markov limit. The most important non-Markovian corrections are
dominated by $G(t)$ and the noise terms contained in the O-operator of
Eq.~(\ref{Oop}), they become more significant in the case of smaller $\gamma$
(or longer $\tau$). As a consequence of long environmental memory, the dynamics
of $\la J_z\ra$ has a long ''tail",  which means that it needs more time to
reach its final steady state compared with the case of the Markov limit
(Fig.~1(d)). The Markov dynamics emerges when $\tau$ becomes shorter and
shorter, that is, for the large $\gamma\gg 1$, $G(t)$ and the noise terms can be
effectively neglected.

As a measure of degree of decoherence, we now consider the purity dynamics of
the three-level system as shown in Fig.~2. It can be shown that the purity
varies from $1$ for a pure state to $1/d$ ($d$ is the dimension of the density
matrix, here we have $d=3$) for a maximally mixed state. We begin with a pure
initial state $|\psi_0\ra=1/\sqrt{3}(|0\ra+|1\ra+|2\ra)$, so for the
zero-temperature case, the final state of the three-level system is also pure.
As shown in Fig.~2, the decoherence pathways of the three-level system are
profoundly modified by the environmental memory parameter $\gamma$. An
interesting feature of the quantum trajectory is that it can reveal how a
quantum state evolves into decoherence with only a small number of
realizations. For example, Fig.~2(a) shows the results generated by only $5$
realizations. Clearly, one cannot expect that the $5$-realization simulations
can quantitatively reproduce an accurate description about decoherence
dynamics, nevertheless, they still give rise to a rather interesting
qualitative picture about the general dissipative behaviors of the purity
including the correct information about the decay and recovery tendencies. For
the numerical simulations with $1000$ realizations shown in Fig.~2(b), it can
be shown that a reliable physical picture can be obtained. Specifically, when
$\gamma=0.2$, the system exhibits a stronger non-Markovian feature, as such,
the exchanges of quantum information (the energy distribution among the three
levels $|0\ra, |1\ra, |2\ra$ and the coherence between any two levels) with the
environment via dissipation proceed slowly and information dissipated into
environment may come back to the system in a finite time. Consequently, the
decoherence time is effectively prolonged. On the contrary, for short memory
time with $\gamma=2$, as shown in Fig.~2(b), the system quickly evolves into a
mixed state and then relaxes itself to the final pure state. When $\gamma=0.8$,
the purity dynamics shows a moderate non-Markovian behavior.

In conclusion, we present an exact non-Markovian quantum state diffusion
equation for the dissipative three-level model described by Eq.~(\ref{Hamil}).
We instigated significant and important progress by obtaining an explicit form
of the O-operator at zero temperature. With the time-local non-Markovian QSD
equation, we are able to attack the transient property of quantum decoherence
dynamics of the three-level system in all possible non-Markovian regimes and
the well-known Markov limit. There are many important connections of our
current work with multiple qubit and qutrit systems, a more detailed research
into this subject will be useful.

We acknowledge discussions with Dr. Wei-Min Zhang. This research
is supported by grants from DARPA QuEST HR0011-09-1-0008, the NSF
PHY-0925174, and NSFC (10804069).

\end{document}